\documentstyle[aps,preprint,epsfig]{revtex}

\begin{document}


\tightenlines


\title{Dipole scattering in polarization induced two-dimensional electron gases}

\author{Debdeep Jena \thanks{Electronic Mail: djena@indy.ece.ucsb.edu}, 
Arthur C. Gossard, Umesh. K. Mishra}
\address{Department of Electrical and Computer Engineering, University of California,
Santa Barbara, CA 93106}
\author{}
\address{}

\maketitle

\begin{abstract}

Unusually large spontaneous and piezoelectric fields in the III-V nitrides have led to 
the making of an entirely new class of two-dimensional electron gas.  Fluctuation from 
a perfectly periodic binary structure in highly polar semiconductor alloys present the same 
physical situation as a random distribution of microscopic dipoles.  The excess dipole 
distribution in the barrier layers is evaluated by a method similar to the virtual crystal 
approximation.  It is shown that the mobility of electrons in the two-dimensional electron 
gas formed in highly polar heterostructures is intrinsically limited by scattering from such 
dipoles.

\end{abstract}



\section{Introduction}

The last decade has witnessed a formulation of the microscopic quantum theory of polarization 
based on a rigorous definition of polarization in a periodic system [Ref. 1-3].  The study
of polarization and its effects was limited to ferroelectrics before the advent of III-V
nitride semiconductors.  The spontaneous and piezoelectric polarization fields of
this class of semiconductors is unusually high compared to other III-Vs; in fact, it 
has given birth to a new frontier of semiconductor 
physics.  The wide band gap and strong polarization fields have found wide application in high 
power, high speed electronic and optical devices [Ref. 4,5].  A large amount of effort has been 
devoted to the 
fabrication and study of high power AlGaN/GaN heterostructure high electron mobility transistors 
(HEMTs) [Ref. 6].  Of chief importance in such devices is the electron mobility in the 
two-dimensional electron gas (2DEG) formed at the heterointerface.  The strong polarization of 
the III-V nitride heterostructures makes a clear analysis of the effect of microscopic 
polarization on mobility of considerable interest to device designers and theorists alike.

At high temperatures ($T\geq 100K$) polar optical phonon scattering is the largest scattering
mechanism; the effects of the other scattering mechanisms become dominant only at low
temperatures.  The set of scattering mechanisms needed to understand the low temperature 2DEG 
mobility of the AlGaAs/GaAs and Si-metal oxide field effect transistor (MOSFET)
is fairly complete.  Theoretical mobility calculations show very good 
agreement with experimentally observed values [Ref 7,8].  However, the 2DEGs in III-V nitride MDHs have 
a fundamentally new origin.  The 2DEG in such heterostructures can be {\em entirely} polarization 
induced, as 
opposed to by remote doping as in AlGaAs/GaAs MDHs and gate-induced inversion in Si-MOSFETs.  
In fact the spontaneous and piezoelectric polarization is large enough to produce 2DEGs without 
intentionally doping the barrier, leading to the novel concept of piezoelectric doping in such 
systems [Ref. 9,10].  The term `modulation doped hetersostructures' (MDHs) is somewhat erroneous 
in such devices, so we prefer to call it the more general HEMTs.   

Previously published calculations of mobilities in AlGaN/GaN 2DEGs have restricted
their analysis to the set of scattering mechanisms that exists for AlGaAs/GaAs 2DEGs [Ref. 11,12].  
This existing set of well understood scattering mechanisms is insufficient for polarization 
induced 2DEGs.  It needs to be expanded by inclusion of the effects of the 
different origins of polarization induced 2DEGs.  This work is directed towards this extention
for the fundamentally new 2DEGs.  In section II, we
theoretically derive the transport scattering rates for 2DEG electrons due to dipoles.  In 
section III, the results are applied to study the III-V nitride heterostructure 2DEG mobilities.
In Section IV, we conclude with a discussion of the results , evaluation of the implications of 
the newly identified scattering mechanism and suggestion of a method to reduce it's effect.

\section{Theory}

Scattering by dipoles and their effects on electron transport in bulk semiconductor samples 
has been studied, albeit not extensively owing to it's insignificance in the non-polar
Si and relatively weakly polar GaAs material systems [Ref. 13,14].  However, the effect of dipole 
scattering on 2DEG electron transport has not been studied to the best of our knowledge.  
We derive the scattering rate due to dipoles for a semiconductor two dimensional electron gas.   

We consider the 2DEG to be perfect (i.e. ,the extent along the $z$ direction to be zero) 
for our derivation.  Extention to the more physical case of a 2DEG with finite extent 
along the growth direction involves incorporation of the relevant form factors.  

Figure[1] shows the model for the system under consideration.  The dipole charges are separated 
from each other
by distance $d_{0}$, and the center is a distance $z$ from the plane containing the 2DEG.  
Spontaneous and piezoelectric polarization fields (${\bf P_{sp}}$ and ${\bf P_{pz}}$ 
respectively) in wurtzite AlGaN/GaN is directed perpendicular to the 2DEG plane [Ref. 15].  
We choose the direction of the dipole to be perpendicular to the 2DEG plane to reflect this. 

The unscreened 
Coulomb potential at the origin due the dipole is written as

\begin{equation}
V_{uns}(r,z)= \frac{e}{4 \pi \epsilon_{0} \epsilon_{b}} \cdot [ 
		\frac{e}{\sqrt{r^2 + (z-\frac{d_{0}}{2})^2}} -
			\frac{e}{\sqrt{r^2 + (z+\frac{d_{0}}{2})^2}}].
\label{simpeq}
\end{equation}

Here $e$ is the electron charge, $\epsilon_{0}$ is the permittivity of free space, 
$\epsilon_{b}$ is the dielectric constant of the subsrate semiconductor, and $r$ is
the in-plane radius vector. 

We evaluate the Fourier transform of this potential in the wavevector ($q$) space    
$V_{uns}(q)= \int V_{uns}(r) e^{i {\bf q}\cdot {\bf r}} d^{3}r $
to get

\begin{equation}
V_{uns}(q,z)= \frac{e^2}{2 \epsilon_{0} \epsilon_{b}} \cdot
		\frac{2e^{-qz} sinh(\frac{qd_{0}}{2})}{q},
\label{simpeq}
\end{equation}

where $q$ is the $x-y$ in-plane wavevector. 


The Fourier component of screened and unscreened potentials are related for a degenerate
2DEG through the Thomas Fermi approximation of the Lindhard formula for the momentum dependent
static dielectric constant [Ref. 16].  It is written as $\epsilon_{2D}(q)=1+\frac{q_{TF}}{q}$, 
where $q_{TF}$ is the Thomas-Fermi wavevector, defined as $\frac{2}{a_{B}^{*}}$, $a_{B}^{*}$ 
being the effective Bohr radius in the semiconductor containing the 2DEG.  A valley degeneracy 
of one and spin degeneracy of two is implied.  The relation is

\begin{equation}
V_{scr}(q,z)=\frac{V_{uns}(q,z)}{\epsilon_{2D}(q)}=\frac{e^2}{2 \epsilon_{0} \epsilon_{b}} 
\cdot
		\frac{2e^{-qz} sinh(\frac{qd_{0}}{2})}{q+q_{TF}}.
\label{simpeq}
\end{equation}

This is the final screened potential experienced by an electron in the 2DEG due to a 
{\em single} dipole at a distance $z$ from the 2DEG plane.

Scattering rate from a state $|{\bf k}>$ to a state $|{\bf k+q}>$ is now evaluated.
Born approximation holds good for evaluation of matrix elements for dilute dipole concentrations. 
Transport scattering rate by a dilute perturbing potential in the Born approximation is written
as [Ref. 17]
 
\begin{equation}
\frac{1}{\tau_{tr}^{2D}}=n_{imp}^{2D} \frac{2 \pi}{\hbar} 
	\int \frac{q^2}{2k^2} |V_{scr}^{tot}(q)|^2 \delta [E(k+q)-E(k)] \frac{d^2 q}{(2 \pi)^2},
\label{simpeq}
\end{equation}

where $n_{imp}^{2D}$ is the 2D impurity concentration, $k$ is the 2D electron wavevector
before scattering, $E(k)$ and $E(k+q)$ are the electron energies before and after scattering,
the $\delta$ function is a statement of the elastic nature of the scattering, and
for our case, $n_{imp}^{2D}$ is the density of dipoles at each Al(Ga) plane, $n_{dipole}^{2D}$.  
$V_{scr}^{tot}(q)$ is the total screened potential experienced by the 2DEG due to the entire
distribution of dipoles.  

Figure [2] illustrates the physical location of the dipoles in AlGaN/GaN HEMTs.  Their origin
is discussed in section III.  Owing to the 
interface roughness, there are dipoles located at the interface too; however, their effect on 
the 2DEG mobility was found to be much less than the far denser distribution of dipoles in the 
barrier.  We consider the 2DEG to be physically located at the centroid of the spatially 
extending quasi-2DEG for illustrating the role of dipoles.  A Fang-Howard wavefunction approach 
would yield respective multiplicative form factors, and is a simple extention of the theory 
presented here.

The screened potential due to the distribution of dipoles in the barrier is hence given by a
Fourier-weighted sum over all dipoles [Ref. 16]

\begin{equation}
V_{scr}^{tot}(q)=\sum_{i} e^{i{\bf q} \cdot {\bf r_{i}}} 
				\frac{V_{uns}(q,z_{i})}{\epsilon_{2D}(q)}.
\label{simpeq}
\end{equation}

If we assume that the dipole distrubution on each Al(Ga) plane are completely uncorrelated,
the cross terms arising in the sum cancel, and we are left with a sum over different 
planes.  The complex exponential can then be factored out and therefore does not contribute 
to the matrix element.  For a thick AlGaN barrier, this evaluates to 

\begin{equation}
V_{scr}^{tot}(q)= \frac{e^{2}}{2 \epsilon_{0} \epsilon_{b}} \cdot 
		\frac{2e^{-q(z_{0}+c_{0})}}{1-e^{-qc_{0}}} \cdot
		\frac{sinh(\frac{qd_{0}}{2})}{q+q_{TF}},
\label{simpeq}
\end{equation}
where $z_{0}$ is the distance of the centroid of the 2DEG from the interface [Figure [2]], and
$c_{0}$ is the separation of the planes containing the dipoles in the barrier.

For a degenerate gas as in a 2DEG, scattering takes place mainly among electrons with 
wavevectors near the Fermi wavevector $k_{F}= \sqrt{2 \pi n_{s}}$ where $n_{s}$ is the 2DEG 
sheet density of carriers.  So the $k$ in the integral can be replaced by $k_{F}$. The elastic 
nature of the scattering process leads to a relation $q=2k_{F} \cdot sin(\frac{\theta}{2})$ 
where $\theta$ is the angle between the wavevectors ${\bf k}$ and ${\bf k+q}$.  Using these 
facts, the expression for scattering rate simplifies to

\begin{equation}
\frac{1}{\tau_{dipole}^{2D}}=n_{dipole}^{2D} \frac{m^{*}}{2 \pi \hbar^{3} k_{F}^{3}} 
	\int_{0}^{2k_{F}} |V_{scr}^{tot}(q)|^2 \frac{q^2 dq}{\sqrt{1-(\frac{q}{2k_{F}})^2}}.
\label{simpeq}
\end{equation}

Using the screened scattering potential of the distribution of dipoles developed in Equation [6]
in Equation [7], we get the final scattering rate due to dipoles.




\section{Application to III-V nitride HEMTs}

We concentrate on AlGaN/GaN heterostructures in our analysis.  Recent work shows that the
2DEG in such heterostructures is created by the subtle interplay of spontaneous and
piezoelectric polarization [Ref. 15,18,19].  The {\em ab initio} calculations of Bernardini 
{\em et al} show that the III-V nitrides have unusually large polarization coefficients 
(an order of
magnitude larger than AlGaAs/GaAs systems) [Ref. 15].  The 2DEG in such heterostructures
can be entirely polarization induced.  In that sense, it is fundamentally different from the 
2DEG in 
AlGaAs/GaAs heterostructures, which is got by modulation doping.  It is argued that the
mobility in such a 2DEG should be affected by the cause of its very existence.

For the perfectly periodic III-V nitride crystal, the microscopic picture of polarization is
a dipole in each primitive cell aligned along the (0001) axis.  The dipole moment 
${\bf p_{0}}=e \cdot d_{0}$ ($d_{0}$ is the effective charge separation) is related to the 
macroscopic polarization ${\bf P}$ by the relation 
${\bf P} = {\bf p_{0}}/{\Omega}$, where $\Omega$ is the volume of the primitive cell [Ref. 20].
${\bf P}$ is the total polarization, which includes the spontaneous and piezoelectric 
components.

\begin{equation}
{\bf P}={\bf P_{sp} + P_{pz}}
\end{equation} 
      
A perfect binary polar lattice thus has a periodically arranged array of dipoles with equal
dipole moments.  Such a periodic arrangement of similar dipoles has a characteristic 
wavevector, and hence does not contribute to the scattering matrix element.

However, the 2DEG in AlGaN/GaN heterostructures is confined by a barrier due to the undoped 
$Al_{x}Ga_{1-x}N$ ternary alloy barrier.  The alloy is a disordered system with Al and Ga atoms
arranged in a random array such that the overall composition over any plane is constant
over Al(Ga) planes.  The difference in spontaneous and piezoelectric polarizations between AlN 
and GaN implies that we have a dipole moment of randomly fluctuating magnitude in the barrier.
We adopt a method similar
to the treatment of disordered alloys by virtual crystal approximation to treat dipoles in
disordered polar semiconductor alloys.

We first arrive at the dipole moments in a primitive cell of coherently strained AlN and GaN 
binary 
wurtzite crystals.  ${\bf P_{sp}}$ and the piezoelectric constants for both the semiconductors 
were 
calculated by Bernardini et. al. in their recent paper [Ref. 15].  The piezoelectric 
field in a binary wurtzite primitive cell coherently strained to a $x-y$ lattice constant 
$a(x)$ from it's unstrained lattice constant $a_{0}$ and $c(x)$ from $c_{0}$ in the $z$ direction
is given by the relation [Ref. 18]

\begin{equation}
P_{pz}(x)=2 \cdot \big( \frac{a(x)-a_{0}}{a_{0}} \big) \cdot [e_{31}-e_{33}\frac{C_{13}}{C_{33}}],
\end{equation}  

where $e_{31}$ and $e_{33}$ are the piezoelectric coefficients and $C_{13}$ and $C_{33}$ are
the elastic constants of the crystal structure.  The volume of the primitive hexagonal cell is

\begin{equation}
\Omega(x)= \frac{\sqrt{3}}{2} c_{0}(x) \cdot a_{0}^{2}(x).
\end{equation}

Thus the dipole moment in a strained binary crystal is given by 

\begin{equation}
p_{dipole}(x)=\big( P_{sp}+P_{pz}(x) \big) \cdot \Omega(x).
\end{equation}

This dipole moment is calculated for both semiconductors as $p_{dipole,AlN}(x)$ and 
$p_{dipole,GaN}(x)$.

We model the disordered $Al_{x}Ga_{1-x}N$ barrier as a perfect crystal superposed with 
a randomly fluctuating dipole moment at each primitive cell.  Such a virtual crystal 
has a dipole moment of magnitude 

\begin{equation}
p_{dipole}(av)=x \cdot p_{dipole,AlN}(x) + (1-x) \cdot p_{dipole,GaN}(x).
\end{equation}

The deviation from the perfect virtual crystal at all Al sites is 
$(1-x) \cdot \Delta p_{dipole}$ where 

\begin{equation}
\Delta p_{dipole}=p_{dipole,AlN}(x)-p_{dipole,GaN}(x).
\end{equation}

The deviation at Ga sites is $(-x) \cdot \Delta p_{dipole}$.  Since there are $x$ Al sites
and $(1-x)$ Ga sites on average on a Al(Ga) plane, the average randomly fluctuating dipole 
moment at each site is

\begin{equation}
\delta p_{dipole}= e\cdot d_{0}=2 \cdot x \cdot (1-x) \cdot |\Delta p_{dipole}|.
\end{equation} 

The absolute value is used in adding the dipole contributions since the 
direction of the dipole is immaterial in the scattering matrix element, which involves the 
square of the dipole potential.  The number of such dipoles present on each Al(Ga) plane is
given by 

\begin{equation}
n_{dipole}^{2D}=\frac{1}{\frac{\sqrt{3}}{4}a_{0}^{2}(x)},
\end{equation}

where the in plane lattice constant $a_{0}(x)$ is interpolated for the alloy.

The polarization induced 2DEG sheet density is given by the difference in 
polarization at the interface as a function of alloy fraction $x$ [Ref. 18],

\begin{equation}
n_{s}(x)=|P_{pz}(Al_{x}Ga_{1-x}N)+P_{sp}(Al_{x}Ga_{1-x}N)-P_{sp}(GaN)|.
\end{equation}

Note that the 2DEG concentration does not involve any modulation dopants; it is formed entirely
to satisfy the discontinuity in the polarization fields at the interface.   

\section{Results and discussion}

The mobility inhibited by dipole scattering alone 
$\mu_{dipole}^{2D}=e \tau_{dipole}^{2D}/ m^{*}$ is evaluated for different alloy compositions.  The results
are plotted in Figure [3].  We also plot the polarization induced 2DEG sheet density in the 
same figure for easy comparision. 
In Figure [4], we plot the dipole scattering inhibited 2DEG sheet conductivity given by 
$G=e \mu_{dipole}^{2D} n_{s}$, and finally, in Figure [5], 
we plot the dipole scattering limited mobility as a function of the 2DEG sheet
density. 

An expected increase in mobility with the increase in the binary nature of the alloy barrier is
seen.
The intrinsic low temperature mobility limit in the $x=0.1$ to $x=0.4$ range (which is 
typical of state of the art III-V nitride HEMTs) is in the $400,000 cm^{2}/V s$ 
to $200,000 cm^{2}/V s$ range.  It is well worth noticing that this is much lower
than the record low temperature mobilities ($\approx 10^{7} cm^{2}/Vs$) of AlGaAs/GaAs 
modulation doped heterostructures, and an order of magnitude higher than the record high
mobilities in AlGaN/GaN HEMTs observed till date ($51,700 cm^{2}/Vs$) [Ref. 19,21].  
It hints at some more severe scattering mechanism(s) that determine the low temperature 
mobility in the III-V nitrides.  

Interface roughness was initially thought to be a 
mobility limiting mechanism owing to a) very high 2DEG sheet densities, and b) trying to fit 
experimental data to the existing set of scattering mechanisms for AlGaAs/GaAs MDHs [Ref. 12,22].  
Recent experiments with double heterostructures point towards the contrary [Ref. 23].  The 
III-V nitrides is a fundamentally new system, and it is necessary that we address issues that 
make the material system so different from the AlGaAs/GaAs systems.    

Dislocation scattering was identified as becoming dominant at high dislocation
densities, which result from the lattice mismatch of the epitaxial GaN layer with the present
substrates of choice - SiC or Sapphire [Ref. 24].  The novel method of lateral epitaxial 
overgrowth
(LEO) is a promising candidate for reducing the density of dislocations in the nitrides 
[Ref. 25]. 
The effects of dipole scattering will be the next hurdle to overcome in pushing the mobilities
higher.  Digital alloy growth is suggested as a technique to reduce the severity of dipole
scattering.  By growing either purely Al or purely Ga layers, we use periodicity to overcome
the scattering originating from the random nature of the alloy.  However, digital alloy growth
suffers from interdiffusion of atoms in the growth process, so dipole scattering cannot be
completely eliminated by this method.

\section*{Acknowledgments}
The authors acknowledge illuminating discussions with S. Keller, I. P. Smorchkova, and 
J. S. Speck.  The authors are grateful to I. P. Smorchkova for sharing her data on the 
latest record high low temperature mobilities in AlGaN/GaN 2DEGs.


\pagebreak

\begin{figure}[h]
\begin{center}
\leavevmode
\epsfxsize=3in 
\epsfysize=3in 
\epsfbox{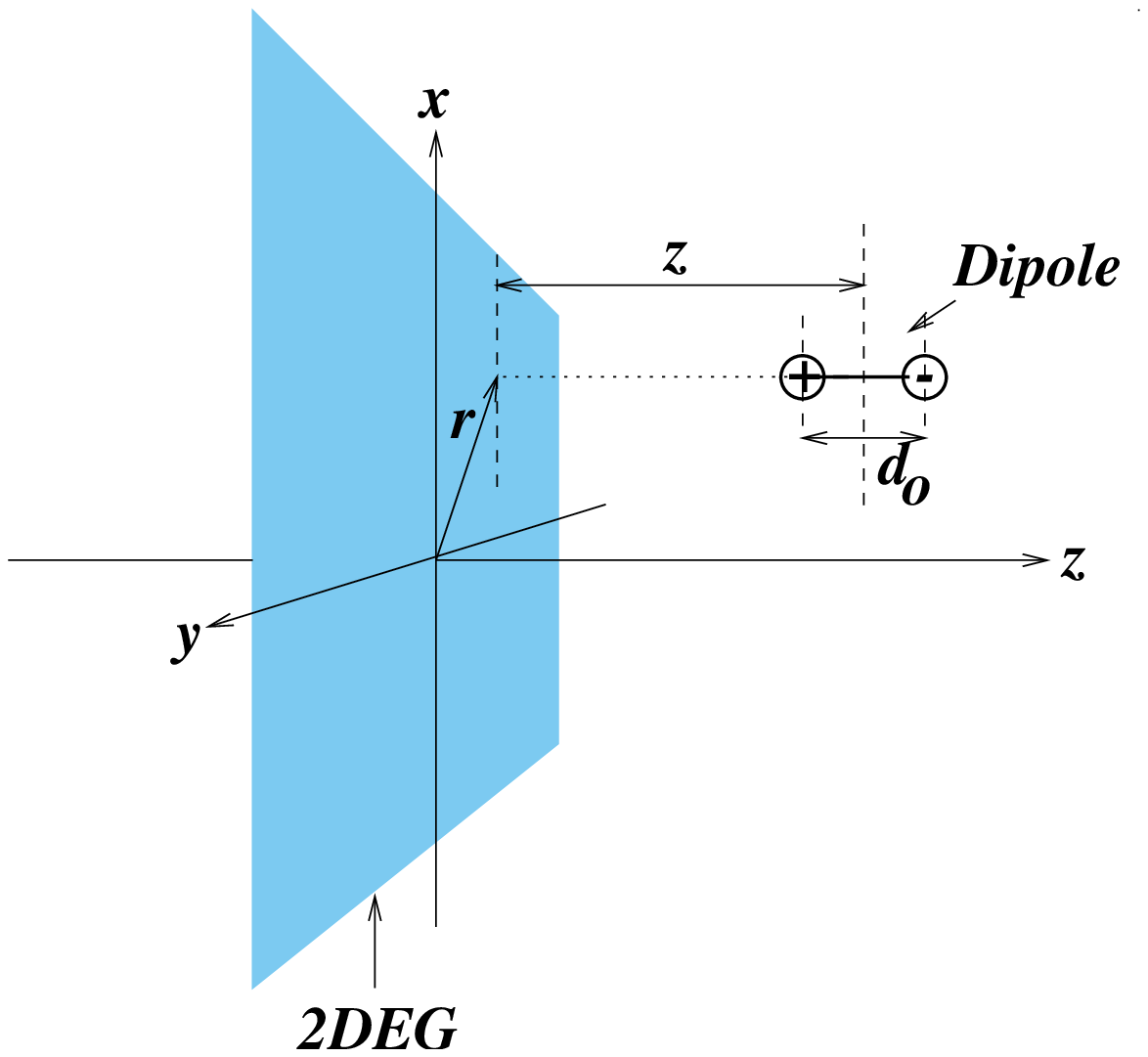}
\end{center}
\caption{The location of the dipole with respect to the 2DEG is shown.  The dipole axis is
taken to be perpendicular to the plane of the 2DEG, keeping with the direction of the 
polarization field in the AlGaN barrier of AlGaN/GaN HEMTs.  The distances used in the 
text in the derivation of the scattering rate are defined.}
\end{figure}

\pagebreak

\begin{figure}[h]
\begin{center}
\leavevmode
\epsfxsize=6in 
\epsfysize=3in 
\epsfbox{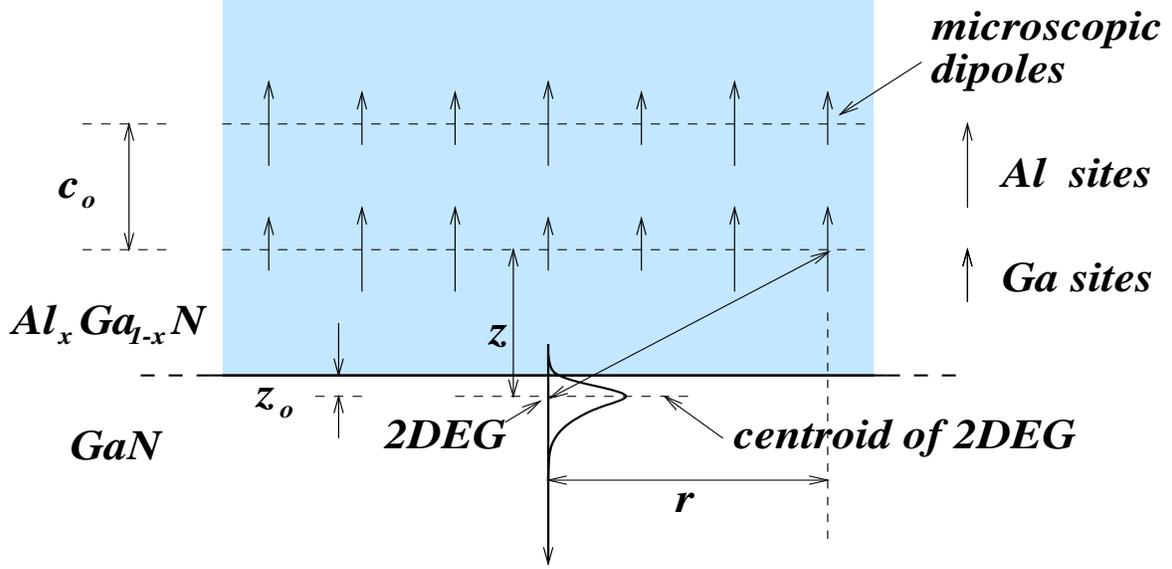}
\end{center}
\caption{The distribution of the dipoles in the AlGaN barrier is shown.  Every Al(Ga) plane
has dipoles in each primitive cell.  The dipole moment at Al sites is higher than that at the
Ga sites owing to the higher spontaneous polarization and piezoelectric constants in AlN than
in GaN.  This fluctuation leads to a random distribution of dipole moments which leads to
scattering of the electrons in the 2DEG.  The 2DEG is assumed to be located entirely at
the centroid of the quasi-2DEG distribution for simplicity.}
\end{figure} 

\pagebreak

\begin{figure}[h]
\begin{center}
\leavevmode
\epsfxsize=4in 
\epsfysize=6in 
\epsfbox{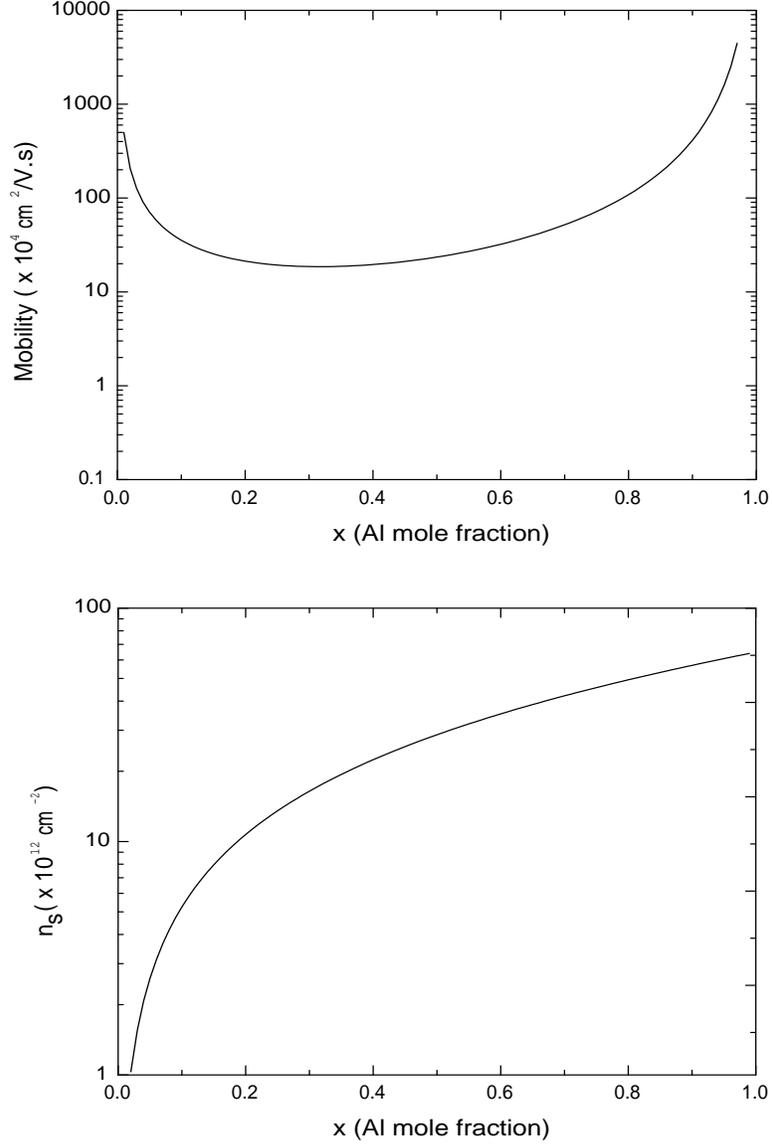}
\end{center}
\caption{Mobility of electrons in the 2DEG inhibited by dipole scattering alone is plotted
as a function of alloy composition.  Sheet density of carriers also changes with alloy 
composition, and is shown in the lower half.  Dipole scattering dominates at alloy compositions
in the $x=0.2-0.5$ range. As the binary nature of the alloy increases, dipole scattering 
reduces, leading to higher mobilities. }
\end{figure} 

\pagebreak

\begin{figure}[h]
\begin{center}
\leavevmode
\epsfxsize=4in 
\epsfysize=3in 
\epsfbox{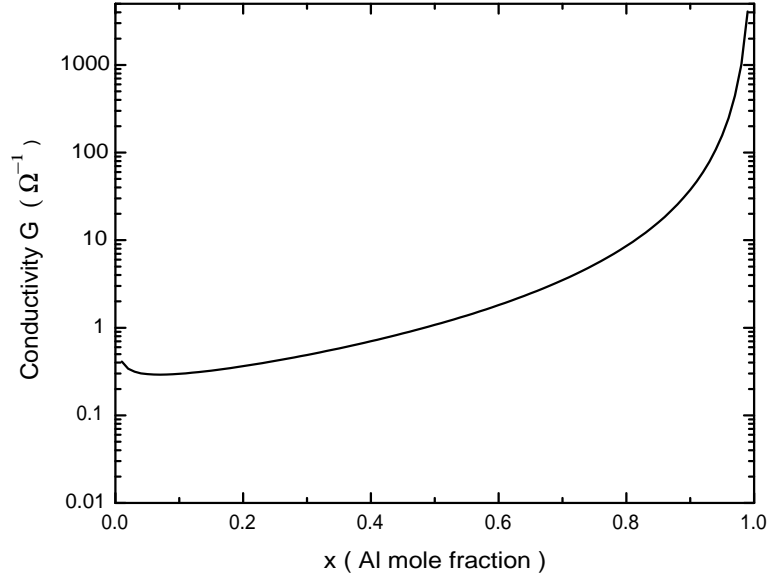}
\end{center}
\caption{Two-dimensional sheet conductivity $G= e \mu_{dipole}^{2D} n_{s}$, 
inhibited by dipole scattering alone is plotted
as a function of alloy composition.  }
\end{figure} 

\pagebreak

\begin{figure}[h]
\begin{center}
\leavevmode
\epsfxsize=4in 
\epsfysize=3in 
\epsfbox{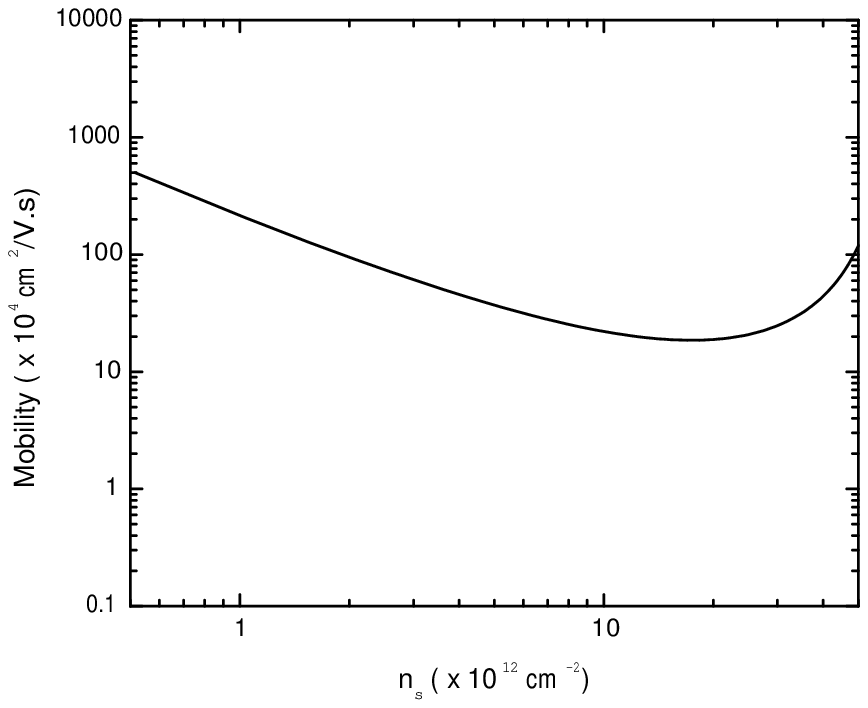}
\end{center}
\caption{Mobility limited by dipole scattering alone is plotted as a function of sheet density
of carriers.  The mobility reaches a minimum at 2DEG sheet density of 
$n_{s} \approx 2 \times 10^{13}/cm^{2}$.}
\end{figure} 

\pagebreak



\begin{references}
\bibitem{ref1} R. D. King-Smith and D. Vanderbilt, Phys. Rev. B {\bf 47}, 1651 (1993).
\bibitem{ref2} R. Resta, Rev. Mod. Phys. {\bf 66}, 899 (1994).
\bibitem{ref3} D. Vanderbilt and R. D. King-Smith, Phys. Rev. B {\bf 48}, 4442 (1993).
\bibitem{ref4} S. Nakamura, M. Senoh, S. I. Nagahama, N. Iwasa, T. Yamada, T. Matsushita,
                         Y. Sugimoto, and H. Kiyoku, Appl. Phys. Lett. {\bf 70}, 1417 (1997).
\bibitem{ref5} H. Sakai, T. Takeuchi, S. Sota, M. Katsuragawa,
                       M. Komori, H. Amano,and I. Akasaki, J. Cryst. Growth {\bf 189/190}, 831 (1998).
\bibitem{ref6} Y. F. Yu, B. P. Keller, P. Fini, S. Keller, T. J. Jenkins, L. T. Kehias, S. P. Denbaars, and U. K. Mishra, IEEE Electron Device Lett. {\bf 19}, 50 (1998).
\bibitem{ref7} W. Walukiewicz, H. E. Ruda, J. Lagowski, and H. C. Gatos, Phys. Rev. B {\bf 30}, 4571 (1984).
\bibitem{ref8} T. Ando, A. B. Fowler, and F. Stern, Rev. Mod. Phys. {\bf 54}, 437 (1982).
\bibitem{ref9} M. A. Khan, J. W. Yang, G. Simin, R. Gaska, M. S. Shur, and A. D. Bykhovski, Appl. Phys. Lett. {\bf 75}, 2806 (1999).
\bibitem{ref10} E. T. Yu, G. J. Sullivan, P. M. Asbeck, C. D. Wang, D. Qiao, and S. S. Lau, Appl. Phys. Lett. {\bf 71}, 2794 (1997).
\bibitem{ref11} L. Hsu and W. Walukiewicz, Phys. Rev. B {\bf 56}, 1520 (1997).
\bibitem{ref12} R. Oberhuber, G. Zandler, and P. Vogl, Appl. Phys. Lett. {\bf 73}, 818 (1998). 
\bibitem{ref13} R. Stratton, J. Phys. Chem. Solids {\bf 23}, 1011 (1962).
\bibitem{ref14} B. K. Ridley, {\em Quantum processes in semiconductors} (Clanderon Press, Oxford, 1982), p.168.
\bibitem{ref15} F. Bernardini, V. Fiorentini, and D. Vanderbilt, Phys. Rev. B {\bf 56}, R10 024 (1997).
\bibitem{ref16} D. K. Ferry and S. M. Goodnick, {\em Transport in Nanostructures} (Cambridge University Press, Cambridge, 1997), p.60-70. 
\bibitem{ref17} J. H. Davies, {\em The Physics of Low Dimensional Semiconductors} (Cambridge University Press, Cambridge, 1998), p.357.
\bibitem{ref18} O. Ambacher, B. Foutz, J. Smart, J. R. Shealy, N. G. Weimann, K. Chu,
                M. Murphy, A. J. Sierakowski, W. J. Schaff, L. F. Eastman, R. Dimitrov, A. Mitchell,
                and M. Stutzmann, J. Appl. Phys. {\bf 87}, 334 (2000).
\bibitem{ref19} I. P. Smorchkova, C. R. Elsass, J. P. Ibbetson, R. Vetury, B. Heying, P. Fini, E. Haus, S. P. Denbaars, J. S. Speck, and U. K. Mishra, J. Appl. Phys. {\bf 86}, 4520 (1999).
\bibitem{ref20} N. W. Ashcroft and N. D. Mermin, {\em Solid State Physics} (Saunders College Publishing, Orlando, 1976), p.555.
\bibitem{ref21} L. N. Pfeiffer, K. W. West, H. L. Stormer, and K. W. Baldwin, Appl. Phys. Lett. {\bf 55}, 1888 (1989).
\bibitem{ref22} Y. Zhang and J. Singh, J. Appl. Phys. {\bf 85}, 587 (1999).
\bibitem{ref23} I. P. Smorchkova (unpublished).
\bibitem{ref24} D. Jena, A. C. Gossard, and U. K. Mishra, Appl. Phys. Lett. {\bf 76}, 1707 (2000).
\bibitem{ref25} T. S. Zheleva, O. H. Nam, M. D. Bremser, and R. F. Davis, Appl. Phys. Lett. {\bf 71}, 2472 (1997).

\end{references}
\end{document}